# Bit Error Rate Prediction of Coded MIMO-OFDM Systems

Y. Nasser *member IEEE*, J.F. Helard *Senior member IEEE,* M. Crussiere *member IEEE*

*Institute of Electronics and Telecommunications of Rennes, UMR CNRS 6464, Rennes, France*

*Email : youssef.nasser@insa-rennes.fr*

## ABSTRACT

Bit error rate (BER) prediction over channel realisations has emerged as an active research area. In this paper, we give analytical signal to interference and noise ratio (SINR) evaluation of MIMO-OFDM systems using an iterative receiver. Using this analytical SINR expression, we propose an accurate BER prediction method based on effective exponential SINR mapping (EESM) method. We show by simulations that our method is independent of the channel realisation and of the MIMO scheme. It is only dependent on the modulation and coding scheme.

*Index Terms-* OFDM, MIMO, Space time block codes, Bit error rate, EESM technique.

## 1. INTRODUCTION

Recently, the combination of multiple input multiple-output (MIMO) and orthogonal frequency division multiplexing (OFDM) techniques is pursued as a potential candidate for the future wireless networks since they ensure high spectrum efficiency as well as high diversity gain. However, the performance of this combination can be further improved if adequate link adaptation algorithms are adopted. The choice of the best algorithm could be based on the modulation and coding schemes (MCS) but also on the applications requirements and channel conditions. As a consequence, an accurate and robust real-time channel prediction is required by the higher layer protocols, in particular bit error rate (BER) prediction [1]. Accurate BER prediction can facilitate design, performance evaluation and parameter tuning of many wireless protocols and applications. For instance, rate-adaptive applications and data link protocols can use accurate BER predictions to adapt their source and channel coding rates in accordance with the forecasted channel conditions.

In this paper, we investigate the effective exponential signal to interference and noise ratio method (EESM) in the MIMO-OFDM systems to predict the BER at the output of the channel decoder. The predicted BER could therefore be used by higher layer protocols in order to adapt their transmission modes. The contribution of this work is twofold. First, a generalized framework is proposed for modelling the signal to interference and noise ratio (SINR) at each iteration of a sub-optimal iterative receiver in MIMO-OFDM systems. Therefore, we adapt the EESM technique, initially validated within 3GPP for OFDM systems [2], to a MIMO-OFDM context. We show that the EESM technique is independent of the channel fading profile and depends only on the MCS. This paper is structured as follows. Section 2 describes the transmission model in MIMO-OFDM systems. In section 3 we give analytical expressions of the SINR at the output of the detector. In section 4 we describe the EESM technique and its adaptation to the MIMO-OFDM system. Section 5 validates our model through simulation results. Conclusions are drawn in section 6.

## 2. TRANSMISSION MODEL

Consider a MIMO-OFDM communication system using $M_T$ transmit antennas (Tx) and $M_R$ receive antennas (Rx). Figure 1 depicts the transmitter modules. Information bits $b_k$ are first channel encoded, randomly interleaved, and fed to a quadrature amplitude modulation (QAM) module which assigns $B$ bits for each of the complex constellation points. Therefore, each group $S=[s_1,…,s_Q]$ of $Q$ complex symbols $s_q$ is fed to a space-time block code (STBC) encoder. Let $X=[x_{i,t}]$ where $x_{i,t}$ ($i=1,…, M_T$; $t=1,…, T$) be the output of STBC encoder. The ST coding rate is then $R = Q/T$. This output is then fed to $M_T$ OFDM modulators, each using $N$ sub-carriers. Then, the signal power at the output of the ST encoder is normalized by $M_T$.

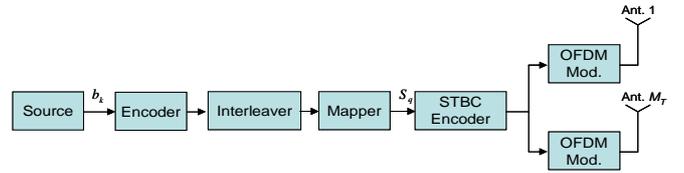

Figure 1- MIMO-OFDM transmitter.

In our transmission model, we assume that the transmitter and receiver are perfectly synchronised. Moreover, we assume perfect channel state information (CSI) at the receiver. Since we assume a frequency domain transmission, the signal received on the sub-carrier $n$ by the antenna $j$ is a superposition of the transmitted signal by the different antennas multiplied by the channel coefficients $h_{j,i}[n]$ to which white Gaussian noise (WGN) is added. It is given by:

$$y_{j,t}[n] = \sum_{i=1}^{M_T} \sqrt{P_i}\, h_{j,i}[n] x_{i,t}[n] + w_{j,t}[n] \qquad (1)$$

where $y_{j,t}[n]$ is the signal received on the $n$th subcarrier by the $j$th receiving antenna during the $t$th OFDM symbol duration. $P_i$ is the assigned power to the symbols transmitted by the $i$th antenna. $h_{j,i}[n]$ is the frequency channel coefficient assumed to be constant during $T$ symbol durations, $x_{i,t}[n]$ is the signal transmitted by the $i$th antenna and $w_{j,t}[n]$ is the additive WGN with zero mean and variance $N_0/2$. In the sequel, we will drop the subcarrier index $n$ for simplicity. By introducing an equivalent receive matrix $\mathbf{Y} \in \mathbb{C}^{M_R \times T}$ whose elements are the complex received symbols expressed in (1) we can write the received signal on the $n^{\text{th}}$ sub-carrier on all receiving antennas:

$$\mathbf{Y} = \mathbf{HPX} + \mathbf{W} \quad (2)$$

where $\mathbf{H}$ is the $(M_R, M_T)$ channel matrix whose components are the coefficients $h_{j,i}$, $\mathbf{P}$ is a $(M_T, M_T)$ diagonal matrix containing the signal magnitudes $\sqrt{P_i}$, $\mathbf{X}$ is a $(M_T, T)$ complex matrix containing the transmitted symbols $x_i[t]$. $\mathbf{W}$ is a $(M_R, T)$ complex matrix corresponding to the WGN.

Let us now describe the transmission link with a general model independently of the ST coding scheme. We separate the real and imaginary parts of the complex symbols input vector $\mathbf{s}$ $\{s_q: q=1,\ldots,Q\}$, of the outputs $\mathbf{X}$ of the ST encoder, and the received signal $\mathbf{Y}$. Let $s_{q,\Re}$ and $s_{q,\Im}$ be the real and imaginary parts of $s_q$. The main parameters of the double code are given by its dispersion matrices $\mathbf{U_q}$ and $\mathbf{V_q}$ corresponding (not equal) to the real and imaginary parts of $\mathbf{X}$ respectively. With these notations, $\mathbf{X}$ is given by:

$$\mathbf{X} = \sum_{q=1}^{Q} \left( s_{q,\Re} \mathbf{U_q} + j s_{q,\Im} \mathbf{V_q} \right) \quad (3)$$

We separate the real and imaginary parts of $\mathbf{S}$, $\mathbf{Y}$ and $\mathbf{X}$ and stack them row-wise in vectors of dimensions $(2Q,1)$, $(2M_R T,1)$ and $(2M_T T,1)$ respectively. We obtain:

$$\begin{aligned}
\mathbf{s} &= \left[ s_{1,\Re}, s_{1,\Im}, \ldots, s_{Q,\Re}, s_{Q,\Im} \right]^{tr} \\
\mathbf{y} &= \left[ y_{1,\Re}, y_{1,\Im}, \ldots, y_{T,\Re}, y_{T,\Im}, \ldots, y_{M_R T,\Re}, y_{M_R T,\Im} \right]^{tr} \\
\mathbf{x} &= \left[ x_{(1,1),\Re}, x_{(1,1),\Im}, \ldots, x_{(2M_T,T),\Re}, x_{(2M_T,T),\Im} \right]^{tr}
\end{aligned} \quad (4)$$

where $tr$ holds for matrix transpose. Since we use linear ST coding, the vector $\mathbf{x}$ can be written as [3]:

$$\mathbf{x} = \mathbf{F}.\mathbf{s} \quad (5)$$

where $\mathbf{F}$ has the dimensions $(2M_T T, 2Q)$ and is obtained through the dispersion matrices of the real and imaginary parts of $\mathbf{X}$ [3].

As we change the formulation of $\mathbf{s}$, $\mathbf{x}$, and $\mathbf{y}$ in (4), it can be shown that vectors $\mathbf{x}$ and $\mathbf{y}$ are related through the matrix $\mathbf{G}$ of dimensions $(2M_R T, 2M_T T)$ such that:

$$\mathbf{y} = \mathbf{GBx} + \mathbf{w} \quad (6)$$

The matrix $\mathbf{B}$ is a $(2M_T T, 2M_T T)$ diagonal matrix whose components are given by:

$$\begin{aligned}
B_{i,i} &= \sqrt{P_i} \quad \text{with } 2.T(p-1)+1 \le i \le 2T.p \\
p &= 1, \ldots, M_T
\end{aligned} \quad (7)$$

Matrix $\mathbf{G}$ is composed of blocks $\mathbf{G_{j,i}}$ $(j=1,\ldots,M_R; i=1,\ldots,M_T)$ each having $(2T,2T)$ elements [3]. Its components are the real and imaginary parts of the channel coefficients. Now, substituting $\mathbf{x}$ from (5) in (6), the relation between $\mathbf{y}$ and $\mathbf{s}$ becomes:

$$\mathbf{y} = \mathbf{GBFs} + \mathbf{w} = \mathbf{G_{eq}} \mathbf{s} + \mathbf{w} \quad (8)$$

$\mathbf{G_{eq}}$ is the equivalent channel matrix of dimensions $(2M_R T, 2M_T T)$ between $\mathbf{s}$ and $\mathbf{y}$. It is assumed to be known perfectly at the receiving side.

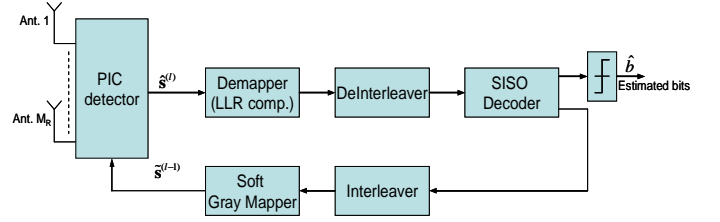

Figure 2- Iterative receiver structure

Now, the detection problem is to find the transmitted data $\mathbf{s}$ given the vector $\mathbf{y}$. The optimal receiver is based on joint ST detection and channel decoding operations. However such receiver is extremely complex to implement and requires large memory for non-orthogonal (NO) STBC codes. Thus the sub-optimal solution proposed here consists of an iterative receiver where the ST detector and channel decoder exchange extrinsic information in an iterative way until the algorithm converges. The iterative detector shown in Figure 2 is composed of a MIMO equalizer, a demapper which is made up of a parallel interference cancellation (PIC) module, a log likelihood ratio (LLR) computation, a soft-input soft-output (SISO) decoder, and a soft mapper. At the first iteration, the demapper takes the estimated symbols $\hat{\mathbf{s}}$, the knowledge of the channel $\mathbf{G_{eq}}$ and the noise variance, and computes the LLR values of each of the coded bits transmitted per channel use. The estimated symbols $\hat{\mathbf{s}}$ are obtained via minimum mean square error (MMSE) filtering according to:

$$\hat{s}_p = \mathbf{g_p}^{tr} \left( \mathbf{G_{eq}} \cdot \mathbf{G_{eq}}^{tr} + \sigma_w^2 \mathbf{I} \right)^{-1} \mathbf{y} \quad (9)$$

where $\mathbf{g_p}$ of dimension $(2M_R T, 1)$ is the $p^{th}$ column of $\mathbf{G_{eq}}$ ($1 \le p \le 2Q$). $\hat{s}_p$ is the estimation of the real part ($p$ odd) or imaginary part ($p$ even) of the complex symbols $s_q$ ($1 \le q \le Q$). The soft Gray mapper takes the soft LLR outputs from the SISO decoder and produces estimation $\tilde{s}_p$ of the transmitted symbol. The estimated symbol $\tilde{s}_p$ belongs to the constellation points set. Its estimation is based on the LLR values and the probability value of each bit of a given constellation point [3]. Once the estimation of the different symbols $\tilde{s}_p$ is achieved by the soft mapper at the first iteration, we use this estimation for the next iterations process. From the second iteration, we perform PIC operation followed by a simple inverse filtering (instead of MMSE filtering at the first iteration):

$$\begin{aligned}
\hat{\mathbf{y}}_\mathbf{p}^{(l)} &= \mathbf{y} - \mathbf{G_{eq,p}} \tilde{\mathbf{s}}_\mathbf{p}^{(l-1)} \\
\hat{s}_p^{(l)} &= \frac{1}{\mathbf{g_p}^{tr} \mathbf{g_p}} \mathbf{g_p}^{tr} \hat{\mathbf{y}}_\mathbf{p}^{(l)}
\end{aligned} \quad (10)$$

where the superscript $(\cdot)^l$ refers to the iteration number. $\mathbf{G_{eq,p}}$ of dimension $(2M_R T, 2Q-1)$ is the matrix $\mathbf{G_{eq}}$ with its $p^{th}$ column removed, $\tilde{\mathbf{s}}_\mathbf{p}$ of dimension $(2Q-1, 1)$ is the vector

$\tilde{\mathbf{s}}$ estimated by the soft mapper with its $p^{th}$ entry removed. In next sections, we will evaluate the iterative detection process through SINR and BER analysis.

## 3. SINR EVALUATION

Without loss of generality, we assume that we are interested by the $p^{th}$ symbol. Using the vector-matrix notation of previous section, the estimated received symbol at the first iteration in (9) could be written in an equivalent form as:

$$\hat{s}_p = I_0^{(1)} + I_1^{(1)} + I_2^{(1)}$$

$$I_0^{(1)} = \mathbf{g_p^{tr}} \left( \mathbf{G_{eq}} \cdot \mathbf{G_{eq}^{tr}} + \sigma_w^2 \mathbf{I} \right)^{-1} \mathbf{g_p} s_p$$

$$I_1^{(1)} = \sum_{\substack{q=1 \\ q \neq p}}^{2Q} \mathbf{g_p^{tr}} \left( \mathbf{G_{eq}} \cdot \mathbf{G_{eq}^{tr}} + \sigma_w^2 \mathbf{I} \right)^{-1} \mathbf{g_q} s_q \quad (11)$$

$$I_2^{(1)} = \mathbf{g_p^{tr}} \left( \mathbf{G_{eq}} \cdot \mathbf{G_{eq}^{tr}} + \sigma_w^2 \mathbf{I} \right)^{-1} \mathbf{w}$$

In (11), $I_0^{(1)}$ is the useful received signal, $I_1^{(1)}$ is the interference signal between different antennas due to the non-orthogonality of the considered STBC. We can easily verify that it is equal to zero for orthogonal STBC schemes. $I_2^{(1)}$ is the colored noise. The complex transmitted data symbols are assumed i.i.d. having zero mean and unit variance (the variance of the real and imaginary parts is equal to ½). Due to this distribution, the SINR expression can be deduced from (11) by:

$$SINR = \frac{E\left\{\left|I_0^{(1)}\right|^2\right\}}{E\left\{\left|I_1^{(1)}\right|^2\right\} + E\left\{\left|I_2^{(1)}\right|^2\right\}} \quad (12)$$

The expectations values in (12) over the random data symbols are given by:

$$E\left\{\left|I_0^{(1)}\right|^2\right\} = \frac{1}{2}\left|\mathbf{g_p^{tr}} \left( \mathbf{G_{eq}} \cdot \mathbf{G_{eq}^{tr}} + \sigma_w^2 \mathbf{I} \right)^{-1} \mathbf{g_p}\right|^2$$

$$E\left\{\left|I_1^{(1)}\right|^2\right\} = \frac{1}{2}\sum_{\substack{q=1 \\ q \neq p}}^{2Q}\left|\mathbf{g_p^{tr}} \left( \mathbf{G_{eq}} \cdot \mathbf{G_{eq}^{tr}} + \sigma_w^2 \mathbf{I} \right)^{-1} \mathbf{g_q}\right|^2 \quad (13)$$

$$E\left\{\left|I_2^{(1)}\right|^2\right\} = \frac{N_0}{2}\left|\mathbf{g_p^{tr}} \left( \mathbf{G_{eq}} \cdot \mathbf{G_{eq}^{tr}} + \sigma_w^2 \mathbf{I} \right)^{-1}\right|^2$$

At the second iteration, the estimated symbol expressed in (11) becomes more complex. It is obtained using (8) and (9) in (10) by:

$$\hat{s}_p^{(2)} = I_0^{(2)} + I_1^{(2)} + I_2^{(2)} \quad \text{where} \quad (14)$$

$$I_0^{(2)} = s_p$$

$$I_1^{(2)} = \sum_{\substack{q=1 \\ q \neq p}}^{2Q} \frac{1}{\mathbf{g_p^{tr} g_p}} \mathbf{g_p^{tr}} \left[ \mathbf{g_q} s_q - \left( \mathbf{G_{eq}} \cdot \mathbf{G_{eq}^{tr}} + \sigma_w^2 \mathbf{I} \right)^{-1} \mathbf{g_q} \tilde{s}_q^{(1)} \right] \quad (15)$$

$$I_2^{(2)} = \frac{1}{\mathbf{g_p^{tr} g_p}} \mathbf{g_p^{tr}} \mathbf{w}$$

For next iterations, it is clear from (15) that the expressions of the estimated received symbol as well as the estimated SINR become more complex. Therefore, some manipulations should be considered to give an analytical expression of the SINR.

Based on the structure of the iterative receiver, we already know that the outputs of the soft Gray mapper are complex symbols which belong to the constellation points. Let $I_t^{(2)} = I_1^{(2)} + I_2^{(2)} = s_p^{(2)} - I_0^{(2)}$ be the total interference power at the second iteration. Then, two cases can be presented at this stage:

- If the estimated symbol $\tilde{s}_p^{(1)}$ at the output of the Gray mapper is equal to the transmitted symbol $s_p$, the useful signal $I_0^{(2)}$ in (15) is such that $I_0^{(2)} = s_p = \tilde{s}_p^{(1)}$ and the total interference signal at the second iteration becomes:

$$I_t^{(2)} = I_1^{(2)} + I_2^{(2)} = s_p^{(2)} - s_p^{(1)} \quad (16)$$

Since $I_1^{(2)}$ and $I_2^{(2)}$ are independent and the complex outputs of the Gray mapper are zero mean with unit variance, the estimated SINR at the second iteration is:

$$SINR^{(2)} = \frac{E\left\{\left|I_0^{(2)}\right|^2\right\}}{E\left\{\left|I_1^{(2)} + I_2^{(2)}\right|^2\right\}} = \frac{1}{2 \cdot E\left\{\left|\hat{s}_p^{(2)} - \tilde{s}_p^{(1)}\right|^2\right\}} \quad (17)$$

where $\hat{s}_p^{(2)}$ is estimated through (10) and $\tilde{s}_p^{(1)}$ is the output of the soft Gray mapper at the first iteration.

- If the estimated symbol $\tilde{s}_p^{(1)}$ at the first iteration is different from the transmitted symbol $s_p$, the difference between the received signals at the first two successive iterations yields by substituting $I_0^{(2)}$ from (15) in (14):

$$\hat{s}_p^{(2)} - \tilde{s}_p^{(1)} = s_p + I_1^{(2)} + I_2^{(2)} - \tilde{s}_p^{(1)} \quad (18)$$

Since $\tilde{s}_p^{(1)}$ is different from $s_p$ in this case, and the different transmitted symbols are i.i.d., we can verify due to the expectation operation that:

$$SINR^{(2)} = \frac{E\left\{\left|I_0^{(2)}\right|^2\right\}}{E\left\{\left|I_1^{(2)} + I_2^{(2)}\right|^2\right\}} = \frac{E\left\{\left|I_0^{(2)}\right|^2\right\}}{E\left\{\left|I_1^{(2)} + I_2^{(2)} + s_p - \tilde{s}_p^{(1)}\right|^2\right\}}$$

$$= \frac{E\left\{\left|I_0\right|^2\right\}}{E\left\{\left|\hat{s}_p^{(2)} - \tilde{s}_p^{(1)}\right|^2\right\}} = \frac{1}{2 \cdot E\left\{\left|\hat{s}_p^{(2)} - \tilde{s}_p^{(1)}\right|^2\right\}} \quad (19)$$

It is clear from the last term of (17) and (19) that the SINR expression at the second iteration is simpler than that of (15). In this case, only the estimated symbols at each iteration are used for SINR estimation i.e. we don't have to compute complex expressions. Also, we can show that (17) and (19) could be generalized for successive iterations. We will now exploit our theoretical SINR model through BER measurements at the output of the channel decoder.

## 4. BER PREDICTION WITH EESM TECHNIQUE

In the previous section, we derived formulas for the estimation of the SINR at each iteration of the detector output. However, it is desirable to evaluate the system level performance after channel decoding in terms of BER. This work is motivated by the practical need of such measures for accurate and realistic evaluation of the system level performance but also for suitable development of link adaptation algorithms such as adaptive modulation and coding, packet scheduling, hybrid-ARQ, etc [4]. Therefore, an accurate relationship between the SINRs obtained at the output of the detector and the BER performance at the output of the channel decoder must be identified.

Let $J$ denote the packet size in complex data symbols. In general, the data symbols in the packets are transmitted over different resource elements (e.g. sub-carriers) and therefore they may experience different propagation and interference conditions. Thus, the data symbols may have different SINR values. Let **SINR** be the vector of $J$ instantaneous SINR received at the output of the detector. The problem of determining an accurate BER prediction method comes back to looking for a relationship

$$P_e = f(\mathbf{SINR}) \qquad (20)$$

where $P_e$ denotes the bit error probability (BEP) and $f$ is the prediction function, which should be invariant with respect to the fading realization, to the multi-path channel model and should be applicable to different MCS in a soft way, i.e., by changing the values of some generic parameters [5]. In the context of a WGN channel, the SINR becomes a SNR and it remains constant over the packet. In this context, a direct relationship $\xi$ exists between the SNR and the error probability

$$P_{e,WGN} = \xi(\mathbf{SNR}) \qquad (21)$$

The function $\xi$ is called the mapping function. It is obtained through theoretical analysis or link level simulation with a WGN channel. In the general context of a fading channel, where the SINR varies, the prediction function $f$ in (20) can be written exactly as a compound function of the WGN function $\xi$ and a compression function $r$ [5]:

$$P_e = \xi \circ r(\mathbf{SINR}) = \xi(SINR_{eff})$$
$$\text{with} \quad SINR_{eff} = r(\mathbf{SINR}) \qquad (22)$$

The function $r$ is referred to as the compression function since its role is to compress the vector **SINR** of $J$ components into one scalar $SINR_{eff}$. The scalar $SINR_{eff}$ is called the *effective SINR* and it is defined as the SINR which would yield the same error probability in an equivalent WGN channel as the associated vector **SINR** in a fading channel. By writing (22), we have merely turned the problem of determining the prediction function $f$ into the problem of determining the compression function $r$.

In an OFDM system, it was concluded that the key issue to accurately determine the appropriate BER after channel decoding is to use the effective SINR in combination with WGN curves. [2] proposes the EESM technique which is based on the Chernoff Union bound [4] to find the effective SINR. The key EESM technique expression relevant to an OFDM system is given by:

$$SINR_{eff} = -\lambda \ln\left(\frac{1}{N}\sum_{n=0}^{N-1}\exp\left(-\frac{SINR[n]}{\lambda}\right)\right) \qquad (23)$$

$SINR[n]$ is the SINR obtained over the $n^{th}$ sub-carrier and $\lambda$ is a unique parameter which must be estimated from the system level simulations for each MCS. It is estimated once by preliminary simulation for each MCS. When the $SINR_{eff}$ is computed, it will be used for BER prediction at the output of the channel decoder with a simple look-up table (LUT) as shown in Figure 3. This LUT gives the BER at the output of the channel decoder as a function of the SNR for a Gaussian channel. It is computed analytically or by simulations. The uniqueness of $\lambda$ for each MCS is derived from the fact that the effective SINR must fulfill the approximate relation

$$BER(\mathbf{SINR}) = BER_{WGN}(SINR_{eff}) \qquad (24)$$

where $BER_{WGN}$ is the BEP for the WGN channel which depends only on the MCS.

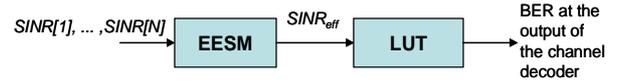

Figure 3- BER prediction through EESM

In our study, the EESM technique must be adapted to a MIMO-OFDM context. Indeed, the estimated received symbol at each sub-carrier is a superposition of different symbols transmitted by the different antennas on that sub-carrier. Therefore, the EESM technique will be applied on the set of Q symbols transmitted on the $M_T$ antennas during $T$ OFDM symbols. The effective SINR is therefore computed through:

$$SINR_{eff} = -\lambda \ln\left(\frac{1}{NQ}\sum_{q=1}^{Q}\sum_{n=1}^{N}\exp\left(-\frac{SINR_q[n]}{\lambda}\right)\right) \qquad (25)$$

## 5. SIMULATION RESULTS

In this section, we validate by simulations our theoretical model based on (12), (19) and (25). In this paper, we consider the orthogonal Alamouti code [5] and the Golden code [7] with $M_T=2$ and $M_R=2$. For equal transmitted powers, we assume that the powers of matrix **B** in (6) are equal to 0 dB. For unequal transmitted powers, we set $P_1$ to 0 dB and we change $P_2$. The considered simulation parameters are given in Table 1. The spectral efficiencies 4 and 6 [b/s/Hz] are obtained for different ST schemes as shown in Table 2. The WGN results are obtained using Alamouti scheme since, NO schemes are not efficient on WGN channel. Figure 4 (resp. Figure 5 ) compares the BER obtained by simulations and the BER predicted with the EESM technique for the Alamouti scheme, considering a spectral efficiency η=4 [b/s/Hz] (resp. η=6 [b/s/Hz]) and different values of transmitted powers. These figures show the accuracy of the proposed technique based on the SINR analytical expression. Moreover, they show that the parameter λ is constant (λ=12.7 for η=4 and λ=22.6 for η=6) and it is independent of the transmitted power but depends on the MCS or equivalently on the spectral efficiency. The parameter λ is obtained by simulations. It is computed once for a given MCS. Figure 6 compares the BER obtained by simulations and the BER predicted with the EESM technique for the Golden code scheme, a spectral efficiency η=6 [b/s/Hz] and different values of transmitted powers. Since the spectral efficiency does not change with respect to results of Figure 5, the parameter λ=22.6 gives an accurate BER prediction and validates our prediction technique.

Table 1- Simulations Parameters

| Number of subcarriers ($N_c$) | 8K mode |
|---|---|
| Guard Interval | 1024 samples |
| Rate $R_c$ of CC | 1/2, 2/3, 3/4 using (133,171)$_o$ |
| Channel estimation | perfect |
| Constellation | 16-QAM, 64-QAM, 256-QAM |
| Spectral Efficiencies | η= 4 and 6 [b/s/Hz] |
| Channel model | Typical Urban (TU-6) |

Table 2- Different MIMO schemes and efficiencies

| Spectral Efficiency | ST scheme | ST rate $R$ | Constellation | $R_c$ |
|---|---|---|---|---|
| η=4 [bit/Sec/Hz] | Alamouti | 1 | 64-QAM | 2/3 |
| | Golden | 2 | 16-QAM | 1/2 |
| η=6 [bit/Sec/Hz] | Alamouti | 1 | 256-QAM | 3/4 |
| | Golden | 2 | 64-QAM | 1/2 |

## 6. CONCLUSION

In this paper, we have proposed an analytical SINR evaluation of MIMO-OFDM systems using iterative receiver. Once the SINR is computed, we have proposed an adaptation of the EESM technique to predict the BER at the output of the channel decoder. We show by simulations the validation of our prediction method. Our future work consists to use this method in cross layer optimization.

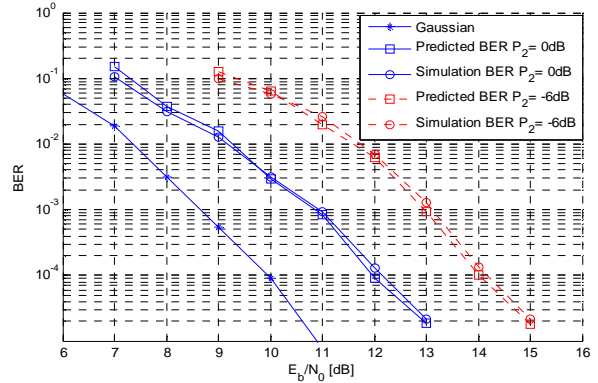

Figure 4- Validation of EESM technique, Alamouti scheme, η=4 [b/s/Hz], λ=12.7

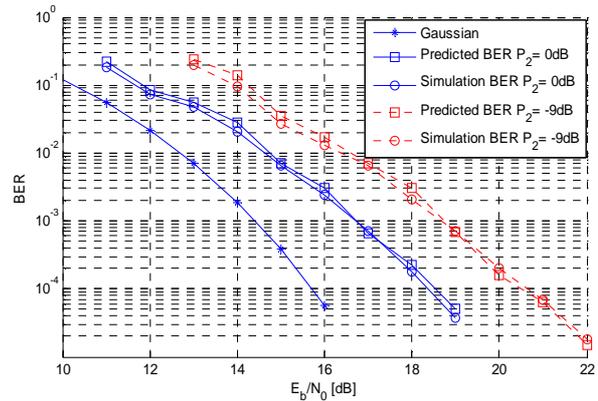

Figure 5- Validation of EESM technique, Alamouti scheme, η=6 [b/s/Hz], λ=22.6

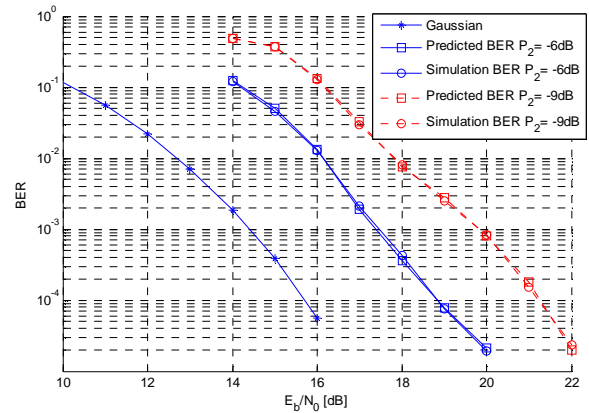

Figure 6- Validation of EESM technique, Golden code scheme, η=6 [b/s/Hz], λ=22.6